\documentclass[aps, prl, amsmath, amssymb, reprint,superscriptaddress, showkeys]{revtex4-2}

\usepackage{graphicx}
\usepackage{xcolor}
\usepackage{mathtools}
\usepackage[normalem]{ulem}
\usepackage{soul}

\renewcommand{\l}{\left}
\renewcommand{\r}{\right}

\begin{document}

\title{
Quantum Enhancement of Thermalization
}

\newcommand{\TUD}{Institut f\"{u}r Theoretische Physik, Technische 
Universit\"{a}t Dresden, D-01062 Dresden, Germany}
\newcommand{\UFPR}{Departamento de F{\'i}sica, Universidade 
Federal do Paran{\'a}, 81531-980 Curitiba, Paran{\'a}, Brazil}
\newcommand{\MPI}{Max-Planck-Institut f\"{u}r Physik komplexer Systeme, N\"othnitzer Stra\ss e 87, D-01187 Dresden, Germany}
\newcommand{\PCS}{Center for Theoretical Physics of Complex Systems, Institute for Basic Science, Daejeon 34126, Korea}
\newcommand{\CESAM}{CESAM research unit, University of Liege, 4000 Liège, Belgium}

\author{Yulong Qiao}
\affiliation{\MPI}
\affiliation{\TUD}
\author{Frank Gro{\ss}mann}
\affiliation{\TUD}
\affiliation{\UFPR}
\author{Peter Schlagheck}
\affiliation{\CESAM}
\author{Gabriel M.~Lando}
\email{gmlando@ibs.re.kr}
\affiliation{\PCS}

\begin{abstract}
Equilibrium properties of many-body systems with a large number of degrees of freedom are generally expected to be described by statistical mechanics. 
Such expectation is closely tied to the observation of thermalization, as manifested through equipartition in time-dependent observables, which takes place both in quantum and classical systems but may look very different in comparison.
By studying the dynamics of individual lattice site populations in ultracold bosonic gases, it is shown that the process of relaxation towards equilibrium in a quantum system can be orders of magnitude faster than in its classical counterpart. 
Classical chaos quantifiers reveal that this is due to a wave packet in a quantum system being able to escape regions of inefficient classical transport by a mechanism akin to tunneling. 
Since the presented phenomenon takes place in a broad parameter range and persists in weakly disordered systems, we expect that it occurs in a variety of many-body systems and is amenable to direct experimental verification in state-of-the-art quantum simulation platforms.
\end{abstract}

\maketitle

Relaxation to equilibrium has proven to be a key tool for studying the thermalization and localization properties of quantum many-body systems \cite{Polkovnikov2011,Trotzky2012,ChoO16S,Kaufman16,Reimann2016,LukO19S,BerO17N}.
State-of-the-art quantum simulation platforms using, e.g., ultracold atoms \cite{Bloch2002} or Rydberg lattices \cite{BerO17N}, allow for the preparation of a non-equilibrium initial state (either directly or via a parameter quench) as well as for the detection of the distribution of particles, spins, or excitations within the system's configuration space after a given evolution time.
An equidistribution found after a sufficiently long time is indicative of eigenstate thermalization \cite{Alessio2016} whereas its absence can often be associated with many-body localization \cite{Abanin19} or, in specific cases, many-body scars \cite{Turner18}.

\begin{figure}[t!]
    \centering
    \includegraphics[width=\columnwidth]{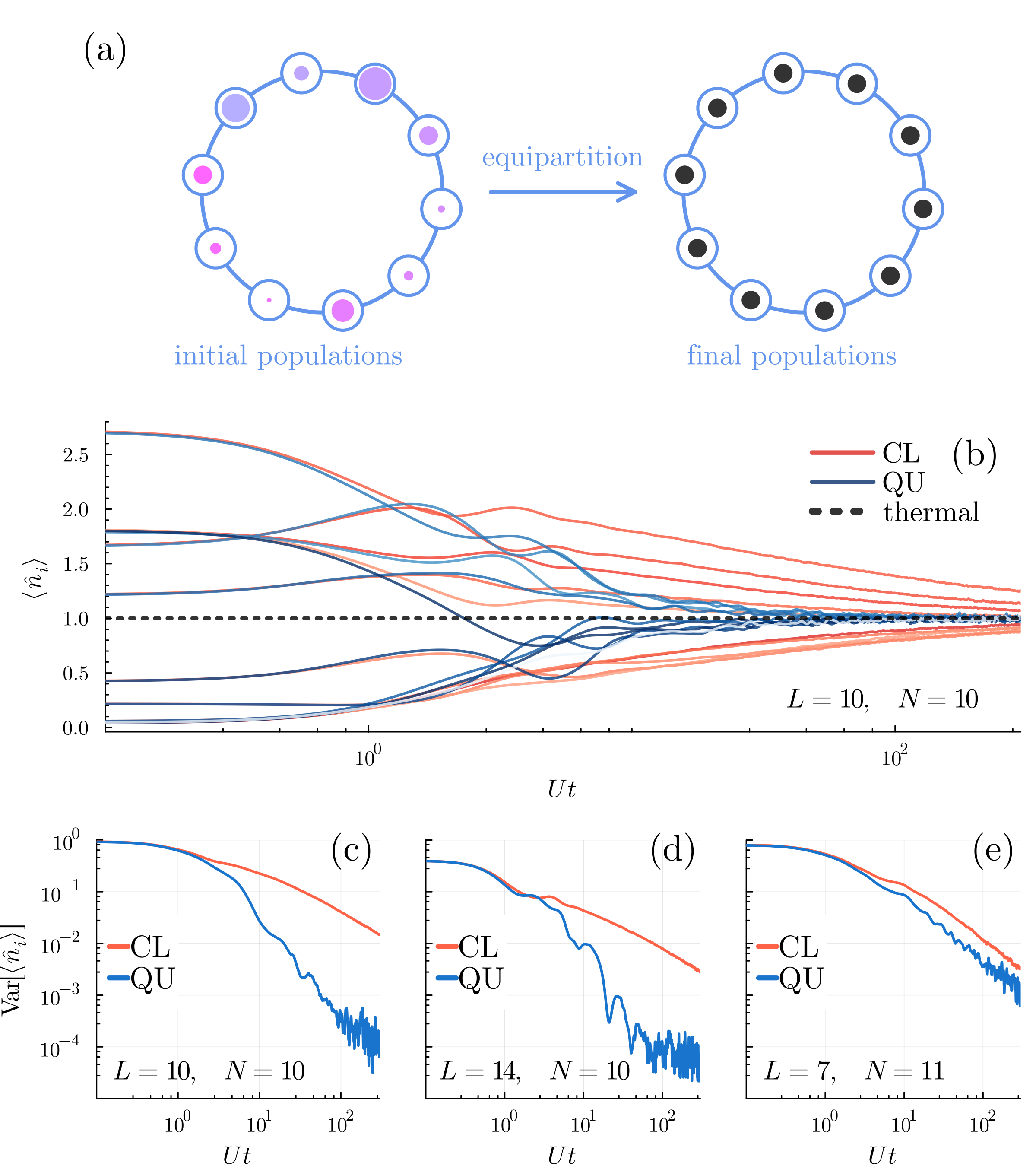}
    \caption{Manifestation of quantum enhancement of thermalization in a Bose-Hubbard ring. (a) A Glauber coherent state with random mean site populations (as indicated by the radii of the filled purple circles) is evolved in the regime of strong interactions, eventually reaching equipartition. (b) Time evolution of individual site populations for $L{=}10$ lattice sites and total lattice population $N{=}10$, calculated quantum mechanically (blue tones) and classically (red tones). They both relax to the thermal value $N/L$ (black dashed line) but on very different time scales. This is most clearly shown in (c) by the time evolution of the corresponding variances with respect to this thermal value, Eq.~\eqref{eq:vars}. Panels (d) and (e) show variances for a larger and a smaller system, which also display faster thermalization of quantum populations when compared to their classical counterparts. All calculations are performed for $J/U{=}0.25$, which corresponds to a regime in which classical dynamics is strongly chaotic (see Fig.~\ref{fig:fig2}).}
    \label{fig:fig1}
\end{figure}

Such relaxation studies have the additional asset that they permit comparisons with the corresponding classical dynamics of the many-body system under consideration, provided the latter exhibits a well-defined and meaningful classical counterpart.
This is, in particular, the case for Bose-Hubbard (BH) systems, which describe ultracold atoms trapped in optical lattices. Their classical analogue, often associated with a mean-field approximation, are Gross-Pitaevskii (GP) lattices. 
Studying such classical analogues provides complementary (and often intuitive) insight into the dynamical transport mechanisms that are at work during evolution, and can be conducted more easily for large system sizes than the corresponding quantum calculations.
Good agreement between classical and quantum dynamics is generally expected for BH systems with relatively weak interaction and/or large site populations, even though localization effects induced by quantum many-body interference can nevertheless give rise to subtle but robust obstacles toward perfect thermalization in the quantum system as opposed to its classical counterpart \cite{richaud2018phase, Engl14}.

As we shall show here, a quantum effect of opposite nature and greater impact is to be encountered in the complementary parameter regime of BH systems, characterized by small onsite populations and a relatively strong interaction as compared to inter-site hopping.
The weakness of the latter inhibits classical transport in the framework of a GP lattice such that thermalization is attained only at very long time scales \cite{livi1987liapunov, Flach09, iubini2019dynamical, lando2023thermalization}. 
For the corresponding quantum system, however, we show here that there are efficient and robust shortcut mechanisms akin to collective many-body tunneling, which bridge phase-space regions of slow transport and thus give rise to a drastic reduction of thermalization time scales as compared to what would be expected from a classical point of view. 
Since such a slowdown in classical diffusion is a property shared by several many-body systems in the weakly coupled regime, we expect this phenomenon to be more general than for BH systems alone.

Specifically, we numerically investigate thermalization in the evolution of Glauber coherent states in BH rings and their classical counterparts. 
We simulate quantum dynamics by means of a multi-configuration variational generalized coherent state approach \cite{Qiao2021, thesisYulong}, while the classical computations are performed using the truncated Wigner approximation \cite{Blakie2008,steel1998dynamical}. 
The efficiency of such methods allows us to reach system sizes large enough to observe relaxation toward thermal equilibrium in the time domain. 
Our results show that while quantum and classical systems behave similarly in the weakly interacting regime, where the onsite interactions are small when compared to hopping, classical equipartition is severely slowed down when compared to its quantum analogue in the complementary strongly interacting case, as showcased in Fig.~\ref{fig:fig1}.
Since equipartition of populations is a reliable indicator of thermalization in this latter regime, this Letter describes the first numerical observation of what can be referred to as the quantum enhancement of thermalization (QET).

We start by explicitly writing the quantum and classical Hamiltonians describing the systems used throughout this Letter. 
BH rings with $L$ sites will be modeled by
\begin{equation}\label{eq:BH_hamiltonian}
    \widehat{H} = \frac{U}{2} \sum_{j=1}^L \widehat{n}_j(\widehat{n}_j - 1) -J \sum_{j=1}^L (\widehat{a}_j^\dagger \widehat{a}_{j+1} + \text{h.c.}) \, ,
\end{equation}
where periodic boundary conditions are assumed and $\widehat{n}_j{=}\widehat{a}_j^\dagger\widehat{a}_j$, with $\widehat{a}_j^\dagger$ ($\widehat{a}_j$) representing bosonic creation (annihilation) operators. The first term above represents onsite interactions, with strength $U$, while the second term models hopping between sites, with strength $J$.
The classical analog of \eqref{eq:BH_hamiltonian} is the GP Hamiltonian 
\begin{equation}\label{eq:GP_hamiltonian}
    H = \frac{U}{2} \sum_{j=1}^L n_j(n_j-1) - J \sum_{j=1}^L \left( p_j p_{j+1}+q_j q_{j+1} \right) \, ,
\end{equation}
where $p_j$ and $q_j$ are the position and momentum quadratures associated to the $j$th degree of freedom and $n_j(p_j,q_j){=}(p_j^2{+}q_j^2{-}1)/2$ is its dequantized number operator. The Hamiltonian in \eqref{eq:GP_hamiltonian} can be obtained from \eqref{eq:BH_hamiltonian} as a mean-field approximation \cite{richter2022semiclassical}. The many-body wave function is then approximated by a classical field known as the order parameter, $\boldsymbol{\psi}\!=\!(\boldsymbol{q}+{\rm i} \boldsymbol{p})/\sqrt{2}$, whose dynamics are described by a discrete nonlinear Schr{\"o}dinger equation. This equation is identical to Hamilton's equations obtained from \eqref{eq:GP_hamiltonian}.

Despite the one-to-one mapping from \eqref{eq:BH_hamiltonian} to \eqref{eq:GP_hamiltonian}, the former describes a quantum system which, unlike the latter, presents several types of phenomena that are absent in classical physics.
These genuinely quantum phenomena can be classified into two main categories depending on whether or not they can still be associated with classical trajectories in the conventional semiclassical framework.
In BH systems such an association would be possible for weak localization in Fock space \cite{Engl14}, for scars \cite{Heller84,Hummel23}, or for many-body revivals \cite{Schlagheck22}, which all arise due to constructive or destructive many-body interference caused by the superposition principle, but are perfectly amenable to quantitative semiclassical reproduction \cite{SimonStrunz,ray2016dynamics,tomsovic2018post,lando2019quantum}.
Phenomena related to barrier or dynamical tunneling \cite{KesSch}, on the other hand, possibly enhanced by chaos and/or resonances \cite{tomsovic1994chaos,brodier2001resonance,brodier2002resonance,Loeck2010,SMU11,Karmakar2021}, can be considered to be intrinsically ``nonclassical'' as no real classical trajectory can describe the associated transition processes in phase space \cite{GrossmannHeller}, and one would have to generalize the semiclassical framework to the complex domain in order to properly incorporate them via complex (e.g., instanton-type) orbits \cite{Miller72,Creagh94,ShudoIkeda95,Takatsuka99,Onishi03}.

The role that such phenomena play in the statistical mechanics of quantum many-body systems is still largely unknown, and in the following we will show that they are responsible for endowing quantum evolution with a much faster approach towards thermalization than purely classical evolution. 
Before moving on, however, it is important to stress that the concept of equipartition, which is employed here to serve as an experimentally accessible indicator for thermalization, does not necessarily imply the presence of the latter under all circumstances, since depending on the choice of the underlying one-body basis (which can be dictated by experimental constraints) equipartition can be observed in the presence of nearly or even fully integrable dynamics \cite{baldovin2021statistical, baldovin2023ergodic}.
For this reason, numerical studies on thermalization have been increasingly focusing on computing Lyapunov spectra or the Kolmogorov-Sinai entropy instead of expectation values \cite{malishava2022lyapunov, lando2023thermalization}, since these are invariant and coordinate-independent quantities that cannot trigger ``false positives'' \cite{eichhorn2001transformation}. 
If picking an observable is unavoidable, a pertinent strategy for choosing it is to focus on functions of the action variables for the nearest integrable limit, simply because right at that limit such actions will be conserved and equipartition cannot take place \cite{arnol2013mathematical, de1988hamiltonian}. 
Thus, if equipartition of such functions persists while approaching integrability and matches statistical mechanics predictions \cite{reimann2013quantum}, it will be necessarily linked to thermalization.

In the context of Hamiltonian \eqref{eq:GP_hamiltonian}, two integrable limits can be identified: the harmonic limit, with $J/U{\to}\infty$; and the limit of decoupled sites, where $J/U{\to}0$.
Due to the near exact quantum-classical correspondence, classical expectation values lie close to their quantum analogues in the neighborhood of the harmonic limit \cite{feynman2010quantum}.
For nearly decoupled sites, however, quantum expectation values will only match their classical counterparts up to Ehrenfest time \cite{schubert2012wave}.
Since the actions in this case are the harmonic oscillator ones, namely $(p_j^2{+}q_j^2)/2{=}n_j{+}1/2$,
single-site populations form the set of ``good'' observables in the 
strongly interacting regime, freezing as $J/U{\to}0$ (but not as $J/U{\to}\infty$, in which case the actions are the normal modes of the lattice). For this reason they will be used to characterize thermalization in the numerical simulations that follow. 

\begin{figure}
    \centering
    \includegraphics[width=\linewidth]{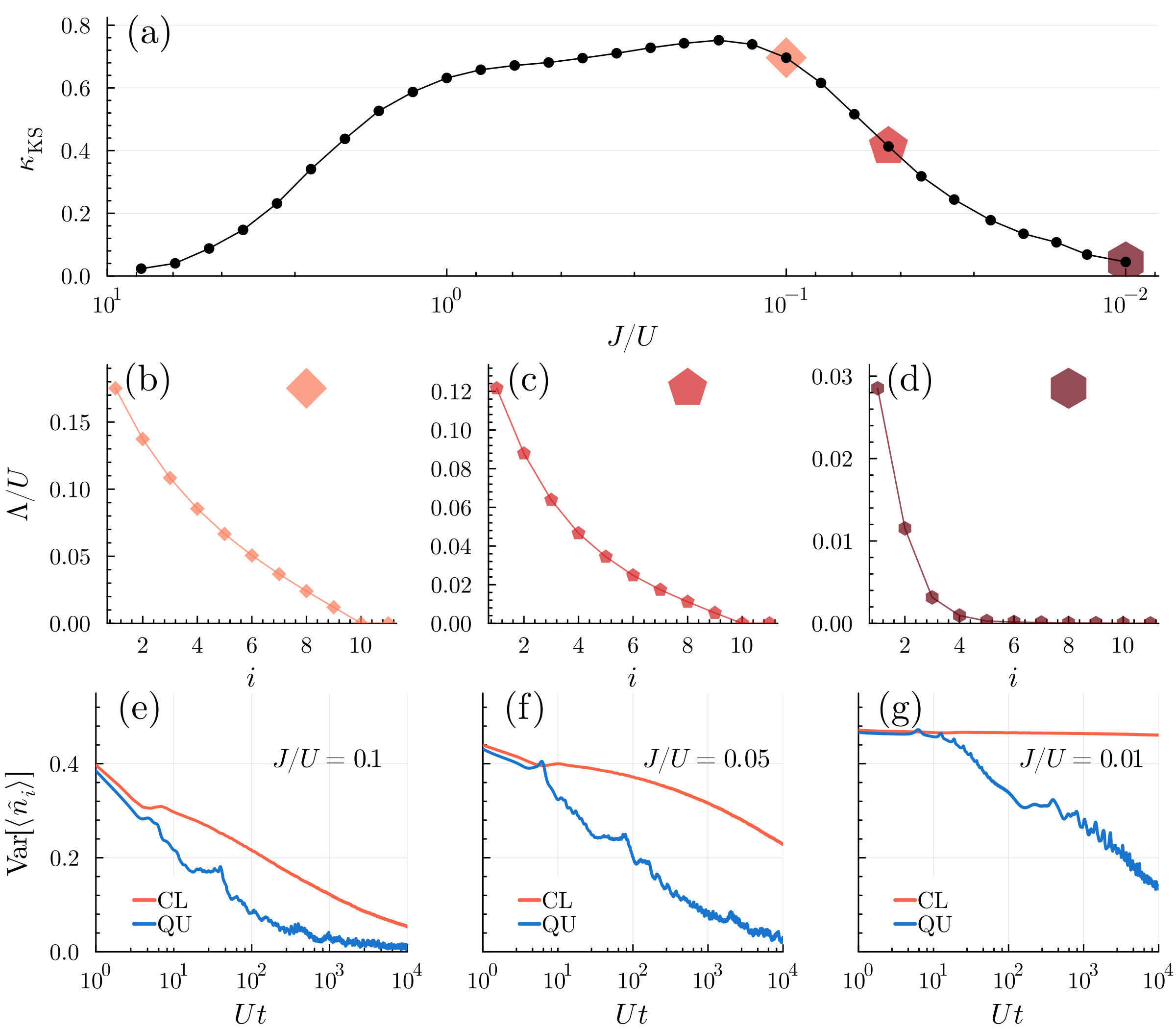}
    \caption{(a) Degree of chaos of the GP lattice as quantified by the Kolmogorov-Sinai entropy, $\kappa_{\mathrm{KS}}$, calculated as a function of decreasing $J/U$ for a randomly chosen initial phase-space point with $N{=}L{=}11$. The colored markers represent regions of maximal (light red, $J/U{=}0.1$), medium (red, $J/U{=}0.05$) and minimal (dark red, $J/U{=}0.01$) chaos, with their corresponding Lyapunov spectra shown in panels (b), (c) and (d) as a function of a counting index, $i$. Panels (e), (f), and (g) show, for the above $J/U$ values, the variances with respect to thermal averages for the quantum BH and the classical GP systems with $L{=}7$ sites calculated starting from Glauber coherent states with random initial populations adding up to $N{=}6$. Note that QET takes place in all (e), (f) and (g) panels, becoming more striking when $J/U$ is decreased.}
    \label{fig:fig2}
\end{figure} 

For an optimal quantum-classical comparison of thermalization we choose initial states given by multi-mode Glauber coherent states,
\begin{equation}\label{eq:CS}
    |n_1, \dots, n_L \rangle = \prod_{j=1}^L \exp \l(  - \frac{n_j^2}{2} + \sqrt{n_j} \, \widehat{a}_j^\dagger \r) | \boldsymbol{0} \rangle \, ,
\end{equation}
where $| \boldsymbol{0} \rangle$ is the vacuum state. 
The coherent state above is the closest quantum analog to a classical phase-space point and is an excellent approximation of Bose-Einstein condensates prepared on single-particle modes with site populations $n_1,\ldots,n_L$ \cite{Lieb2007}, which are often used as initial states in quench experiments \cite{greiner2002collapse}. 
As is well known, the dimension of the Hilbert space for the BH model grows quickly with both the number of lattice sites $L$ and the mean population of the system, $N{=}\sum_j n_j$, rendering many-site calculations  with large $N$ very challenging in Fock space.
We thus use a variational approach, based on a 
multi-layer extension  of the multi-configuration 
ansatz in terms of a linear combination of 
time-dependent generalized coherent 
states with fixed particle number, introduced in \cite{Qiao2021}. Here,
a summation over a range of particle numbers $S$ of the generalized coherent states, 
defined as
$|S,\boldsymbol{\xi}(t)\rangle=\frac{1}{\sqrt{S!}}\left(\sum_{i=1}^L\xi_i(t) \, \widehat{a}_i^\dagger \right)^S|\boldsymbol{0}\rangle$,
is performed in addition to the summation over the ansatz multiplicity,
as explained in detail in \cite{supmat}.

Classical results are obtained via a Monte Carlo evaluation of the expression
\begin{equation}
    \langle \widehat{n}_j \rangle(t) \approx \int_{\mathbb{R}^{2L}} \mathrm{d}\boldsymbol{p} \, \mathrm{d}\boldsymbol{q} \, W_0(\boldsymbol{p}, \boldsymbol{q}) \, n_j \! \left[ \Phi(\boldsymbol{p}, \boldsymbol{q}; t) \right] \, ,
\end{equation}
which is the truncated Wigner approximation for the site populations \cite{steel1998dynamical}. Here, $W_0$ is the Wigner function of the initial state \eqref{eq:CS}, given by a Gaussian in $2L$-dimensional phase space, and $\Phi(\boldsymbol{p},\boldsymbol{q};t)$ denotes the Hamiltonian flow of \eqref{eq:GP_hamiltonian} starting from the initial phase-space point $(\boldsymbol{p}, \boldsymbol{q})$, which is sampled according to $W_0$. 
Sampling over several mean-field trajectories of different energies (we used $10^5{\sim}10^6$ trajectories in our simulations) washes out any behavior associated with atypical solutions, \emph{e.g.}~breathers \cite{flach1998discrete, *rumpf2008transition, *mithun2018weakly}, whose number is much smaller than that of the generic ``thermalizing'' trajectories in the regimes considered here.

Figure~\ref{fig:fig1} demonstrates QET in the regime of strong interactions, with $J/U{=}0.25$.
We consider here a coherent state \eqref{eq:CS} where the populations $n_1,\ldots,n_L$ are chosen as random numbers drawn from a uniform distribution, such that their sum yields the total mean population $N{=}10$ in Fig.~\ref{fig:fig1}(a-d) and $N{=}11$ in Fig.~\ref{fig:fig1}(e).
As can be seen in Fig.~\ref{fig:fig1}(b), quantum and classical time evolutions eventually yield equipartition in the site populations, but on very different time scales: while the quantum calculations reach and stabilize at the thermal average of $N/L$ particles per site at about $t \sim 100/U$, their classical analogues will require times at least one order of magnitude longer in order to reach equilibrium. 
To facilitate comparisons, we express the deviation from thermal equilibrium more quantitatively through the variance
\begin{equation}\label{eq:vars}
    \text{Var} \left[ \langle \widehat{n}_j \rangle(t) \right] = \frac{1}{L-1} \sum_{j=1}^L \left( \langle \widehat{n}_j \rangle(t) - \bar{n}_j \right)^2
\end{equation}
of the site populations with respect to their equilibrium values, namely $\bar{n}_j{=}N / L$ for a perfectly homogeneous lattice.
As is seen in Figs.~\ref{fig:fig1}(c)--(e), QET is most clearly rendered in terms of this variance.
It prevails for large system sizes (Fig.~\ref{fig:fig1}(d)) while it becomes less pronounced for smaller lattices with increased average site populations (Fig.~\ref{fig:fig1}(e) where the system is more semiclassical due to the larger $N/L$ ratio).

How is QET affected by the degree of chaos, which impacts the efficiency of chaotic transport paths towards thermalization in the classical many-body system?
To quantitatively address this question, we show in Fig.~\ref{fig:fig2}(a) the Kolmogorov-Sinai entropy \cite{cassidy2008threshold}, $\kappa_\mathrm{KS}$, for a $L{=}11$ site system with average unit filling, $N/L{=}1$, as a function of $J/U$.
Such entropy is obtained from the sum of positive exponents in the Lyapunov spectrum, computed via the prescription of \cite{benettin1980lyapunov, *geist1990comparison}. 
We choose for this purpose an initial phase-space point with random on-site populations \footnote{We have verified that the Lyapunov spectra and their associated Kolmogorov-Sinai entropies are quantitatively similar for several phase-space points sampled according to the Wigner function of the initial state.}, giving rise to energy densities for which the resulting classical trajectory explores a large domain in phase space ($E/N{\approx}-0.029$ for $J/U{=}0.1$, $E/N{\approx}0.128$ for $J/U{=}0.01$). 
Unsurprisingly, $\kappa_\mathrm{KS}$ vanishes in the harmonic and decoupled integrable limits, while it exhibits a plateau of maximal chaos for $0.1{<}J/U{<}1.0$, in agreement with a previous study on spectral properties of the BH model \cite{pausch2022optimal}. 
QET shown in Fig.~\ref{fig:fig1} is thus occurring at highly developed chaos, far away from any near-integrable limit.

On the strongly interacting side of the plateau, the transition from well-developed chaos to near-integrability is, as is seen in Fig.~\ref{fig:fig2}(b)--(d), manifested in the Lyapunov spectrum by a significant drop in its maximal exponent as well as by a transition from a linear to an exponential decrease of the remaining positive exponents with their index, $i$ \cite{lando2023thermalization}.
The development of near-zero exponents indicates that reducing the coupling strength gives birth to multiple near-conserved quantities.
Correspondingly, as shown in Fig.~\ref{fig:fig2}(e)--(g), chaotic transport in the classical phase space drastically slows down in this regime, eventually giving rise to thermalization time scales that are, for $J/U{=}0.01$, beyond numerical verification.
In the corresponding quantum system, however, this proximity to integrability affects thermalization to a much lesser extent.
Quantum transport in the regime of strong interactions can thus take place along channels that are inaccessible for the corresponding classical system, similarly to tunneling taking place in a truly near-integrable system whose classical phase space is dominantly constituted by invariant Kolmogorov-Arnold-Moser tori.
Note that these quantum channels are already effective at $J/U{\sim}0.1$ where the classical system is maximally chaotic, whereas they do not play any role in the weak interaction regime, which features nearly perfect agreement between the classical and quantum thermalization speeds \cite{supmat}. We note that such regimes of approximate harmonicity or nearly decoupled sites exist in a plethora of models that display the same characteristics in their Lyapunov spectra as seen in Fig.~\ref{fig:fig2}, such as Josephson junction arrays \cite{lando2023thermalization}, in which QET should also take place. 

\begin{figure}
    \centering
    \includegraphics[width=\linewidth]{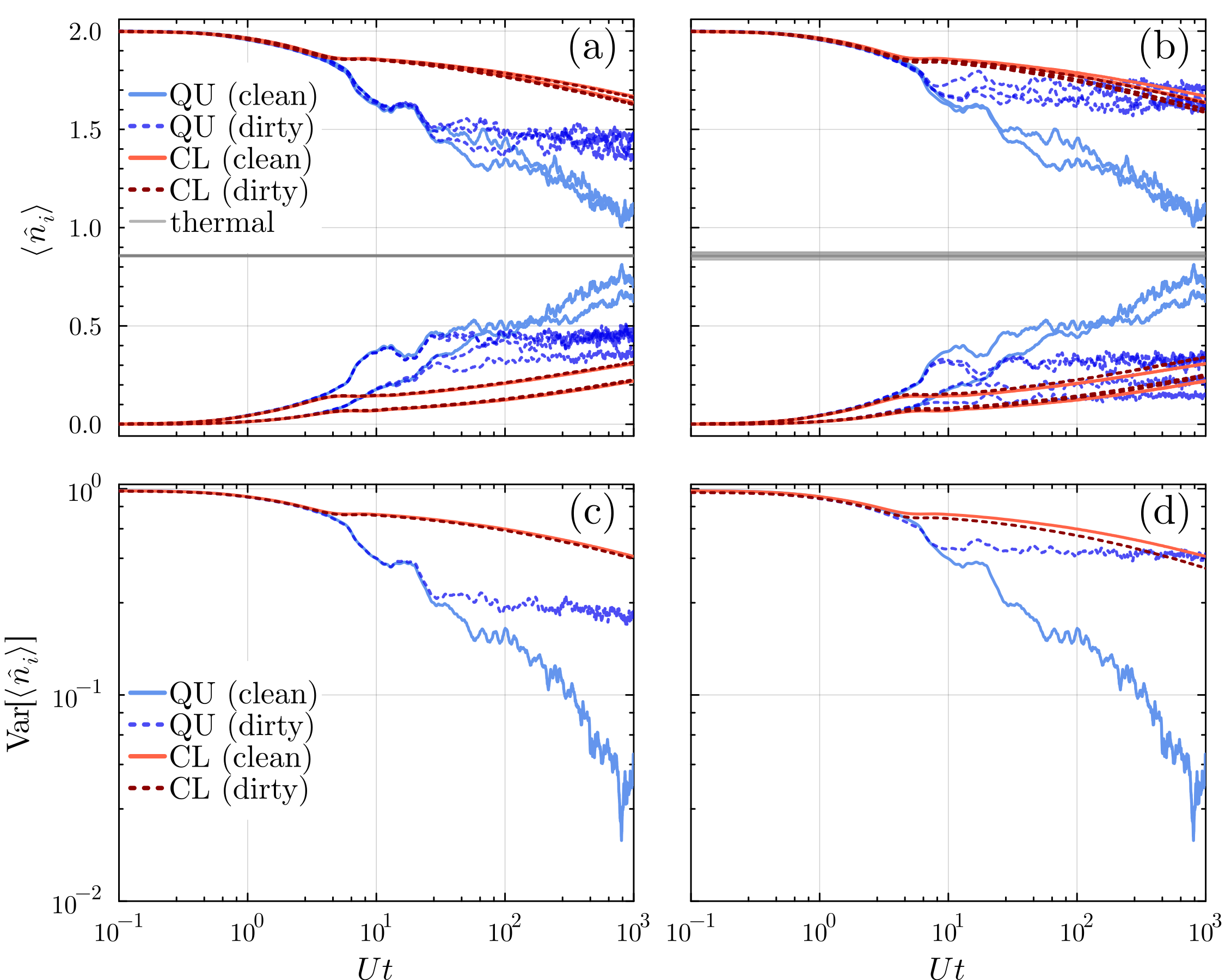}
    \caption{Quantum enhancement of thermalization in the presence of disorder. Shown are quantum and classical mean populations (a,b) and the corresponding total variances with respect to thermal values (c,d) in the absence and presence of disorder (dark and light color tones, respectively), for an initial coherent state with staggered populations $|0,2,0,2,0,2,0\rangle$ within a $L{=}7$ site lattice. Disorder is introduced via a set of randomly chosen onsite energies drawn from a uniform distribution within the interval (a,c) $[-0.05 U, 0.05 U]$,  (b,d) $[-0.25 U, 0.25 U]$.
    While not affecting QET at intermediate time scales the presence of disorder inhibits, on long time scales, equilibration of the site populations to their thermal averages. Since such averages in a disordered system are site dependent, we display all seven of them as gray lines (forming a band) in panels (a) and (b) \cite{supmat}.}
    \label{fig:fig3}
\end{figure}

Despite being of genuine quantum nature, the phenomenon of thermalization enhancement that we encounter here cannot be attributed to the occurrence of resonances in the many-body spectrum caused by accidental near-degeneracies of energy levels.
To demonstrate this, we show in Fig.~\ref{fig:fig3} the time evolution of the mean site populations and their corresponding variance in the presence of weak disorder.
The latter is generated by randomly chosen onsite energies drawn from the uniform intervals $[-0.05U,0.05U]$ (left column) and $[-0.25U,0.25U]$ (right column of Fig.~\ref{fig:fig3}).
Considering, for $L{=}7$ and $J/U{=}0.25$, an initial coherent state with the staggered populations $|0,2,0,2,0,2,0\rangle$ (which in a disorder-free context can give rise to various resonances and near-degeneracies due to its high symmetry), we clearly see that the presence of these weak random energies inhibits equilibration on a long-time scale, even when quantitatively accounting for the fact that the mean thermal site occupancies are no longer uniform for such a specific disorder realization \cite{supmat}.
This slowing down of thermalization is attributed to the disordered system possessing a large number of localized (nonthermal) eigenstates that violate the eigenstate thermalization hypothesis \cite{deutsch1991quantum, srednicki1994chaos}. Dynamical localization in Fock space then hinders the exploration of the system's entire chaotic domain through destructive interference \cite{FishmanGrempelPrange82,Ishikawa2007}.

Rather intriguingly, however, no significant effect of the presence of weak disorder can be found at short and intermediate time scales, $ t {\lesssim} 10/U$, when QET sets in.
We infer from this finding that the enhancement of thermalization is a rather robust phenomenon that does not rely on fine tuning of system parameters.
In spectral terms, while QET cannot be attributed to degeneracies of individual energy levels, we conjecture that it arises from an \emph{approximate} resonance between \emph{groups} of energy levels, namely those that are associated with the occupancy distribution of the initial coherent state and its counterparts obtained through permutations of the occupancies among the lattice sites.
Robustness with respect to system parameter variations is then granted by the multitude of individual site distribution states belonging to these groups, in combination with the fact that the effective coupling between these states can be relatively large, owing to the absence of true tunneling barriers \cite{Loeck2010,SMU11} (as is indeed supported by numerical findings \cite{supmat}).
Hence, even if in the presence of disorder only a few of these states effectively feature an approximate near-degeneracy with components of the initial coherent state, these remaining ``escape channels'' will be sufficient for the quantum system to bridge across slowly diffusive phase-space regions \footnote{Note that in the strong hopping regime this mechanism is expected to become ineffective as compared to transport via classical channels, in close analogy to chaos-assisted tunneling suppressing ordinary tunneling channels between symmetric wells \cite{tomsovic1994chaos}.}.

We expect QET to occur in a variety of physically relevant quantum many-body systems in weakly coupled regimes, ranging from complex molecules and spin chains \cite{bardet2023rapid, kumar2024prethermalization} to quantum computing and simulation \cite{BerO17N,RouO17S}. 
In particular, the thermalization behavior shown in Fig.~\ref{fig:fig2} can be probed in state-of-the-art experimental setups using ultracold bosonic atoms in optical lattices \cite{ChoO16S,Kaufman16,LukO19S}.
Here, the controlled use of environmental effects \cite{witthaut2008,Ott2013} can allow one to destroy many-body quantum coherence and effectively induce classicality (i.e., mean-field dynamics) in the experiment, thus opening possibilities for probing QET through a direct comparison of classical and quantum equipartition speeds.

\acknowledgments

The authors gratefully acknowledge the computing time made available to them on the high-performance computer at the NHR Center of TU Dresden. This center is jointly supported by the Federal Ministry of Education and Research and the state governments participating in the NHR (www.nhr-verein.de/unsere-partner).
GML thanks Sergej Flach and Dominik {\v S}afr{\'a}nek for fruitful discussions, and Technische Universit{\"a}t Dresden for their hospitality. GML acknowledges financial support from the Institute for Basic Science in the Republic of Korea through the project IBS-R024-D1.

\bibliography{bib}

\end{document}


\onecolumngrid

\title{Supplemental material for ``Quantum Enhancement of Thermalization''}

\newcommand{\TUD}{Institut f\"{u}r Theoretische Physik, Technische Universit\"{a}t Dresden, D-01062 Dresden, Germany}
\newcommand{\MPI}{Max-Planck-Institut f\"{u}r Physik Komplexer Systeme, N\"othnitzer Stra\ss e 87, D-01187 Dresden, Germany}
\newcommand{\UFPR}{Departamento de F{\'i}sica, Universidade 
Federal do Paran{\'a}, 81531-980 Curitiba, Paran{\'a}, Brazil}
\newcommand{\PCS}{Center for Theoretical Physics of Complex Systems, Institute for Basic Science, Daejeon 34126, Korea}
\newcommand{\CESAM}{CESAM research unit, University of Liege, 4000 Liège, Belgium}

\author{Yulong Qiao}
\affiliation{\MPI,\TUD}
\author{Frank Gro{\ss}mann}
\affiliation{\TUD,\UFPR}
\author{Peter Schlagheck}
\affiliation{\CESAM}
\author{Gabriel M.~Lando}
\affiliation{\PCS}

\maketitle

\section{Variational multi-configuration ansatz}\label{sec:var}

For the quantum dynamics of the largest systems that we considered, we 
have used a variational multi-configuration approach to solve the 
time-dependent Schr\"odinger equation, based on generalized coherent states 
(GCS), also known as SU($M$) coherent states. A widely employed form 
of  these GCS, as presented in \cite{Buonsante2008,Trimborn2008} is given by
\begin{equation}\label{Eq:GCS}
    |S,\boldsymbol{\xi}\rangle=\frac{1}{\sqrt{S!}}\left(\sum_{i=1}^L\xi_i\widehat a_i^\dag\right)^S|{\boldsymbol{0}}\rangle,
\end{equation}
where $S$ represents the {\it fixed} particle number and 
$\boldsymbol{\xi}=\{\xi_1, \xi_2,\cdots,\xi_L\}$  is a set of complex parameters 
satisfying the normalization condition $\sum_{i=1}^L|\xi_i|^2=1$. 
As in the main text, the symbol $|\boldsymbol{0}\rangle$ denotes the multi-mode vacuum state.
We note that for the choice $\xi_i=1/\sqrt L$, the GCS is the ground state of the free boson model 
(the Bose-Hubbard model with  $U=0$) \cite{DellAnna22}.

In \cite{Qiao2021}, an ansatz in terms of a linear combination 
of GCS with time-dependent amplitudes $A_k(t)$, given by
\begin{equation}\label{multi_GCS}
    |\Psi(t)\rangle=\sum_{k=1}^MA_k(t)|S,\boldsymbol{\xi}_k(t)\rangle,
\end{equation}
was successfully applied to the time-dependent dynamics of the 
Bose-Hubbard model for fixed particle numbers. The 
differential equations for the coefficients 
$\{A_k,\boldsymbol{\xi}_k\}$ were derived using the 
time-dependent variational principle. The accuracy of the 
ansatz depends on the multiplicity $M$. Specifically, $M=1$ 
leads to the discrete nonlinear Schr\"odinger equations for 
the complex  parameters, which leads to the mean-field 
description of the Bose-Hubbard model. Increasing $M$ beyond 
1 allows the result to go beyond the mean-field level and to 
gradually converge to  the exact solution due to the (over-)completeness of the GCS \cite{Qiao2023}.

With the form of the basis function expansion Eq.\ (\ref{multi_GCS}), it is straightforward to calculate expectation values of operators. Given an operator $\hat{O}$, we have
\begin{align}\label{equ:expectation}
    \langle\hat{O}\rangle&=\sum_{k,j}^MA_k^*A_j\langle S,\boldsymbol{\xi}_k(t)|\hat{O}|S,\boldsymbol{\xi}_j(t)\rangle
    =\text{sum}(\boldsymbol{\rho}\circ\boldsymbol{O},\text{all})
\end{align}
where $\boldsymbol{\rho}_{kj}=A_k^*A_j$, and the matrix $\boldsymbol{O}$ has the entries
\begin{align}\boldsymbol{O}_{kj}=\langle S,\boldsymbol{\xi}_k(t)|\hat{O}|S,\boldsymbol{\xi}_j(t)\rangle.
\end{align}
The symbol $\circ$ denotes the Hadamard-product (element-wise 
multiplication of matrices) and ``sum(...,all)'' denotes the summation over all the entries. To make progress, operator ordering is a key concept. In our framework, we consistently apply normal ordering to our operators, ensuring that all creation operators are positioned to the left of the annihilation operators. This systematic arrangement is achieved through the application of the standard bosonic commutation relationship.
If $\hat{O}$ is a normal-ordered operator such as $\hat{a}_f^\dag 
\hat{a}_g^\dag \hat{a}_h^\dag\cdots \hat{a}_l\hat{a}_m\hat{a}_n\cdots$, 
the matrix entries are computed as
\begin{align}
     \boldsymbol{O}_{kj}=S(S-1)\cdots(S-i)\xi_{kf}^*\xi_{kg}\xi_{kh}^*\cdots\xi_{jl}\xi_{jm}\xi_{jn}\cdots\langle S-i,\boldsymbol{\xi}_k|S-i,\boldsymbol{\xi}_j\rangle,
\end{align}
where $i$ is the number of creation (annihilation) operators, and 
we have used the property  $\hat{a}_m|S,\boldsymbol{\xi}_j\rangle=\sqrt{S}\xi_{jm}|S-
1,\boldsymbol{\xi}_j\rangle$ of the GCS. For small values of $M$, the numerical evaluation of Eq.\ (\ref{equ:expectation}) can be performed without too much computational effort.

Eq.\ (\ref{multi_GCS}) is an  ansatz, tailored for systems with a definite 
number of particles. However, the initial multi-mode Glauber coherent state 
given in Eq.\ (3) of the main text, exhibits fluctuations in the particle 
number. To address this issue, we refer to the relationship between GCS 
and Glauber coherent states \cite{Buonsante2008}
\begin{align}\label{GCS_glauber}
|\boldsymbol{\alpha}\rangle=\sum_{N=0}^{\infty}\sqrt{P(S)}|S,\boldsymbol{\xi}_0\rangle,
\end{align}
where $P(S)$ follows a Poisson distribution
\begin{equation}\label{poisson}
    P(S)={\rm e}^{-N}\frac{N^{S}}{S!},
\end{equation}
with a mean equal to the {\it average} particle number of the 
multi-mode Glauber coherent state, given by $N=\sum_{i=1}^L n_j$. 
The characteristic parameters are thus related to the average 
site populations via
\begin{equation}
    \boldsymbol{\xi}_0=\frac{1}{\sqrt{N}}\{\sqrt{n_1}, \sqrt{n_2}, \cdots, \sqrt{n_L}\}.
\end{equation}
Eq.\ (\ref{GCS_glauber}) and Eq.\ (\ref{poisson}) indicate that it is only 
necessary to consider a small range of particle numbers [$S_1$, $S_2$] 
within the significant region of the Poisson distribution, ensuring that the 
overlap of the initial state for the numerics and Eq. (\ref{GCS_glauber}), 
given by $\sum_{S=S_1}^{S_2}P(S)$, is very close to 1.\\
As a result, our comprehensive ansatz for the dynamics initiated as a
multi-mode Glauber coherent state is given by
\begin{align}\label{eq:multi_layer_GCS}
|\Psi(t)\rangle=\sum_{S=S_1}^{S_2}\sqrt{P(S)}\sum_{k=1}^NA_k^{(S)}(t)|S,\boldsymbol{\xi}^{(S)}_k(t)\rangle,
\end{align}
which has a multi-layer structure \cite{thesisYulong}. 
In each layer, labeled by a fixed particle number $S$, the dynamics 
starting from $|S,\boldsymbol{\xi}_0\rangle$ is simulated by using the 
ansatz in Eq.\ (\ref{multi_GCS}). Due to the U(1) 
symmetry of the Bose-Hubbard model, the dynamics in each layer is 
independent of that in any other layer and thus the different layers 
can be solved for in parallel.

\begin{figure}[t]
  \centering
  \includegraphics[width=1.8in]{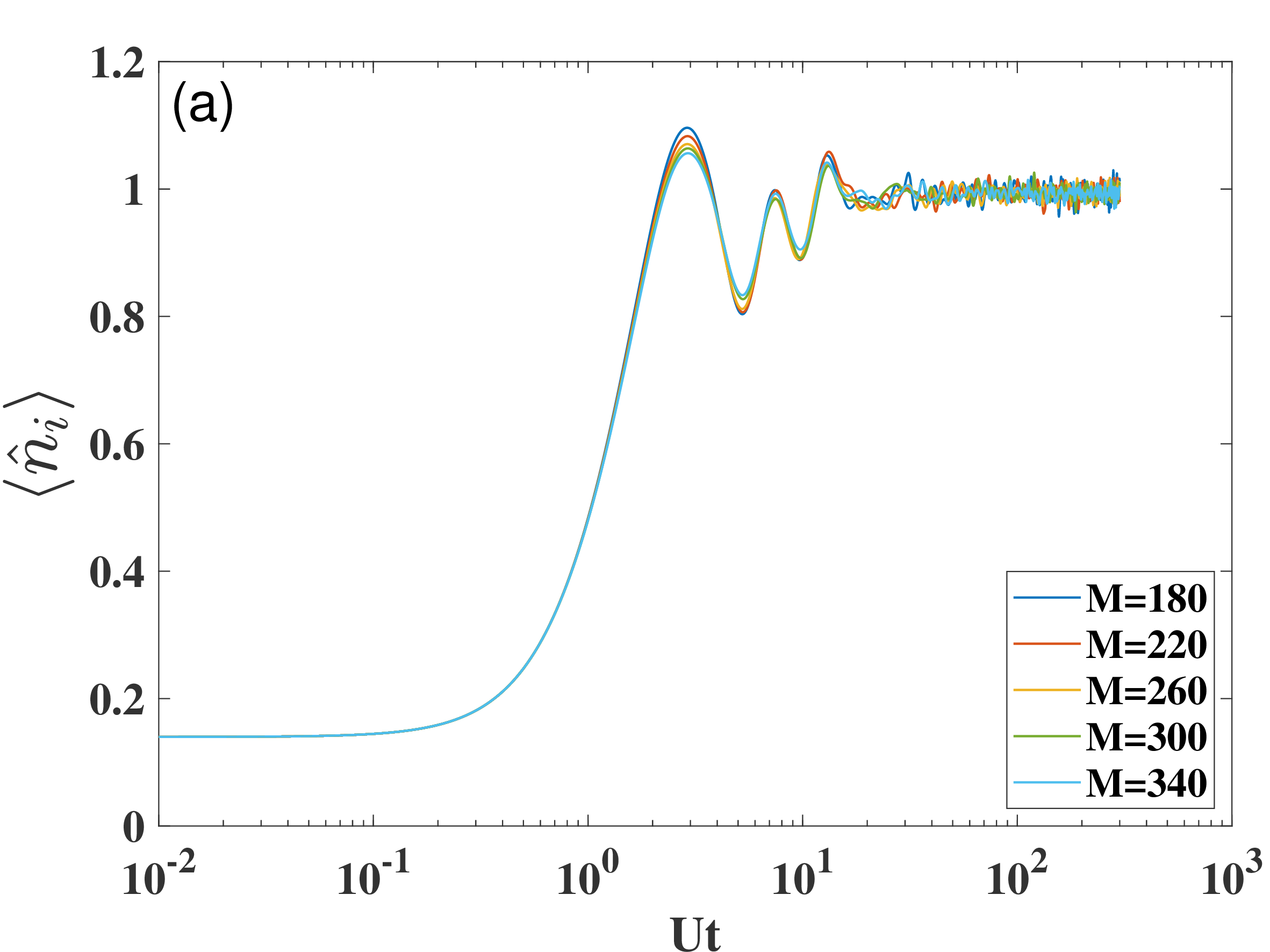}
  \includegraphics[width=1.8in]{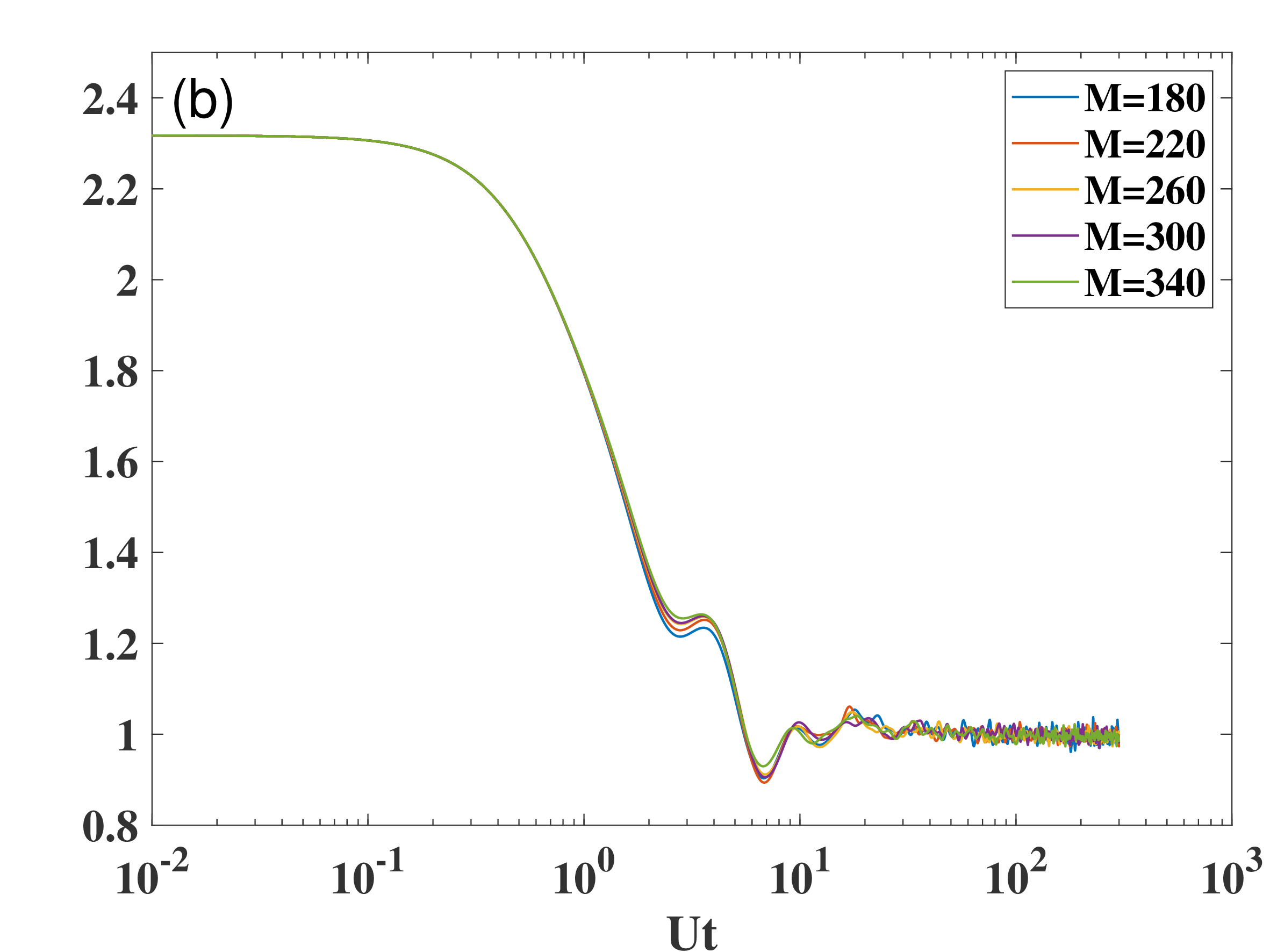}
  \includegraphics[width=1.8in]{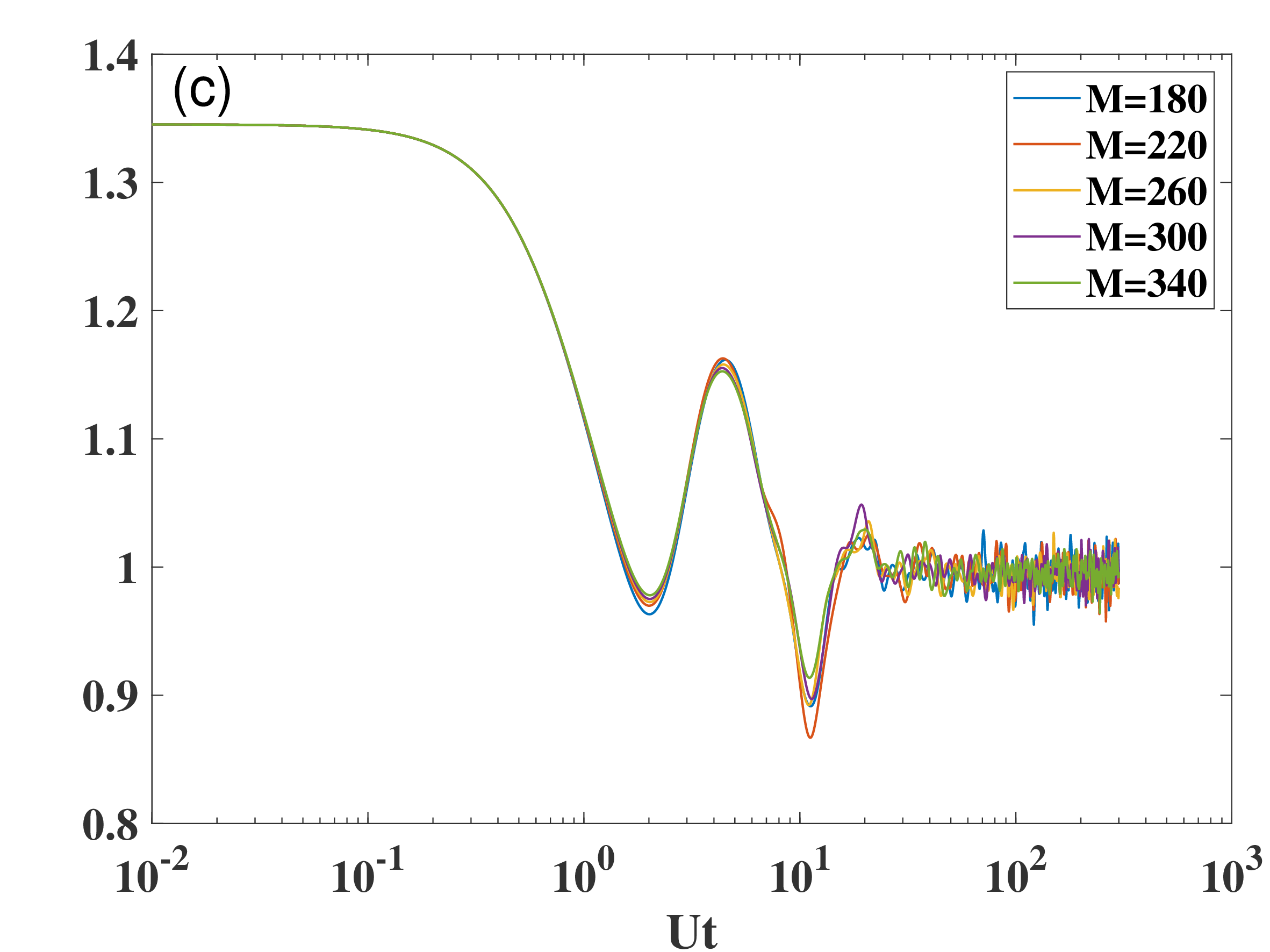}
  \includegraphics[width=1.8in]{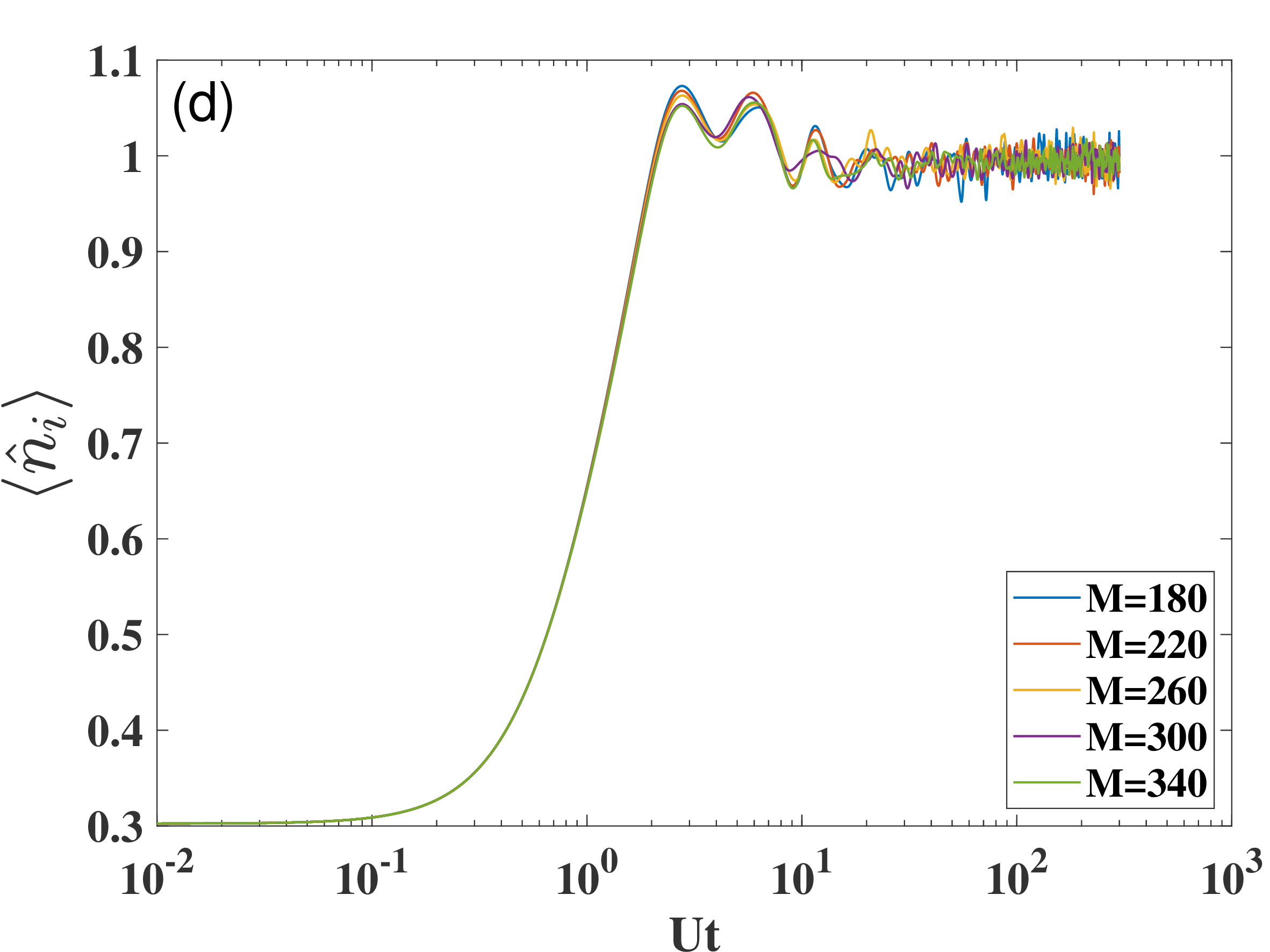}
  \includegraphics[width=1.8in]{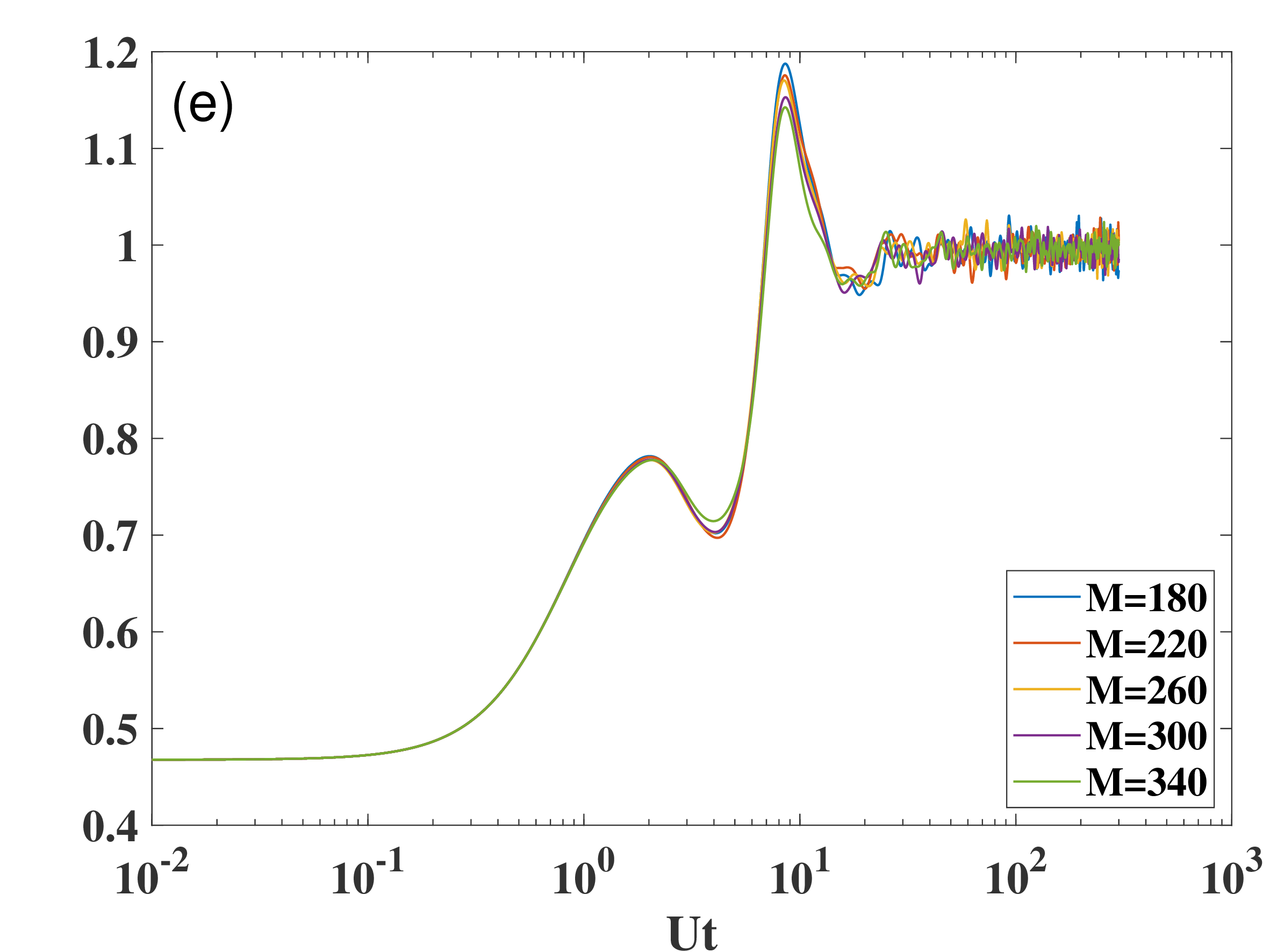}
  \includegraphics[width=1.8in]{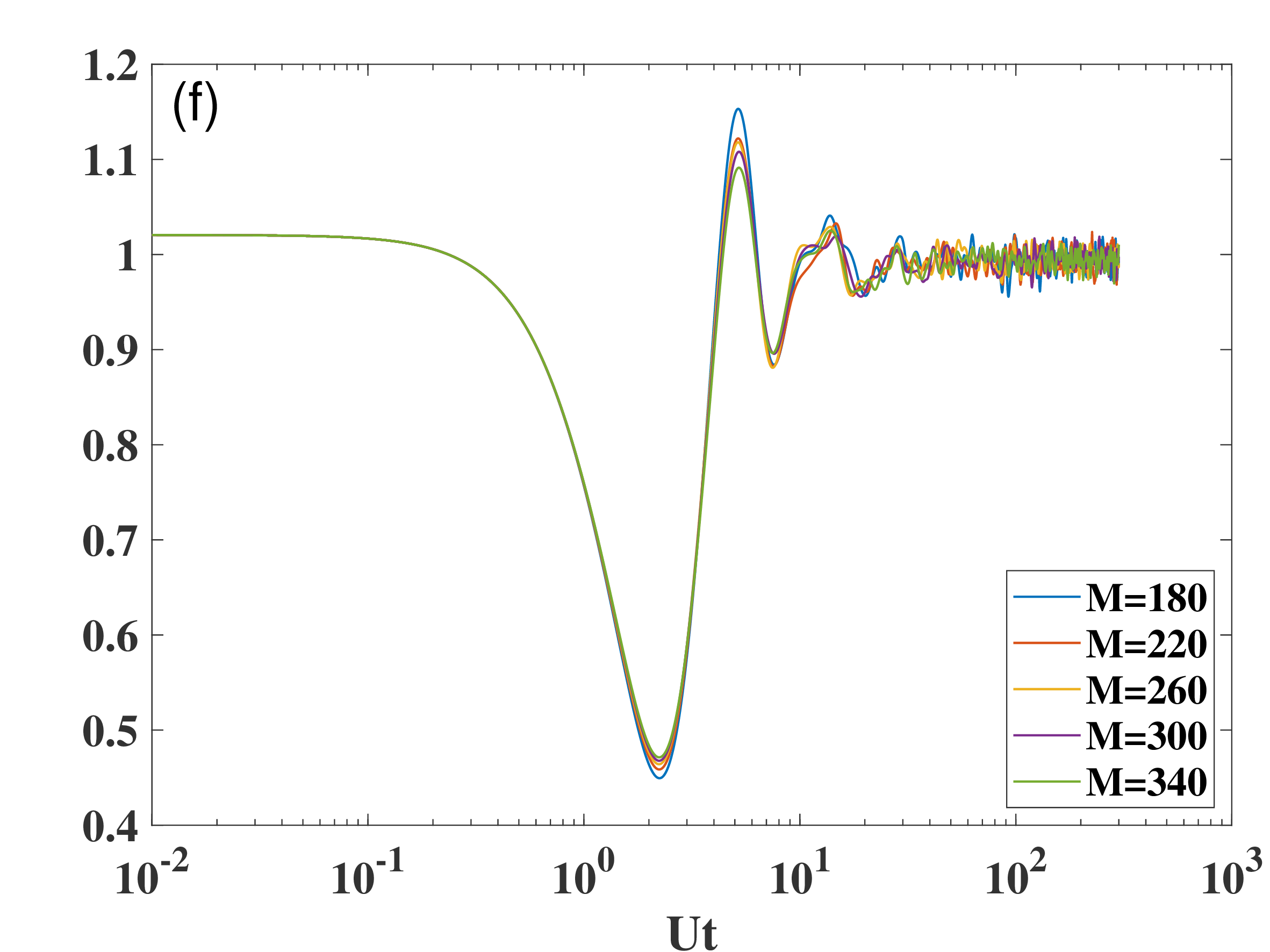}
  \includegraphics[width=5in]{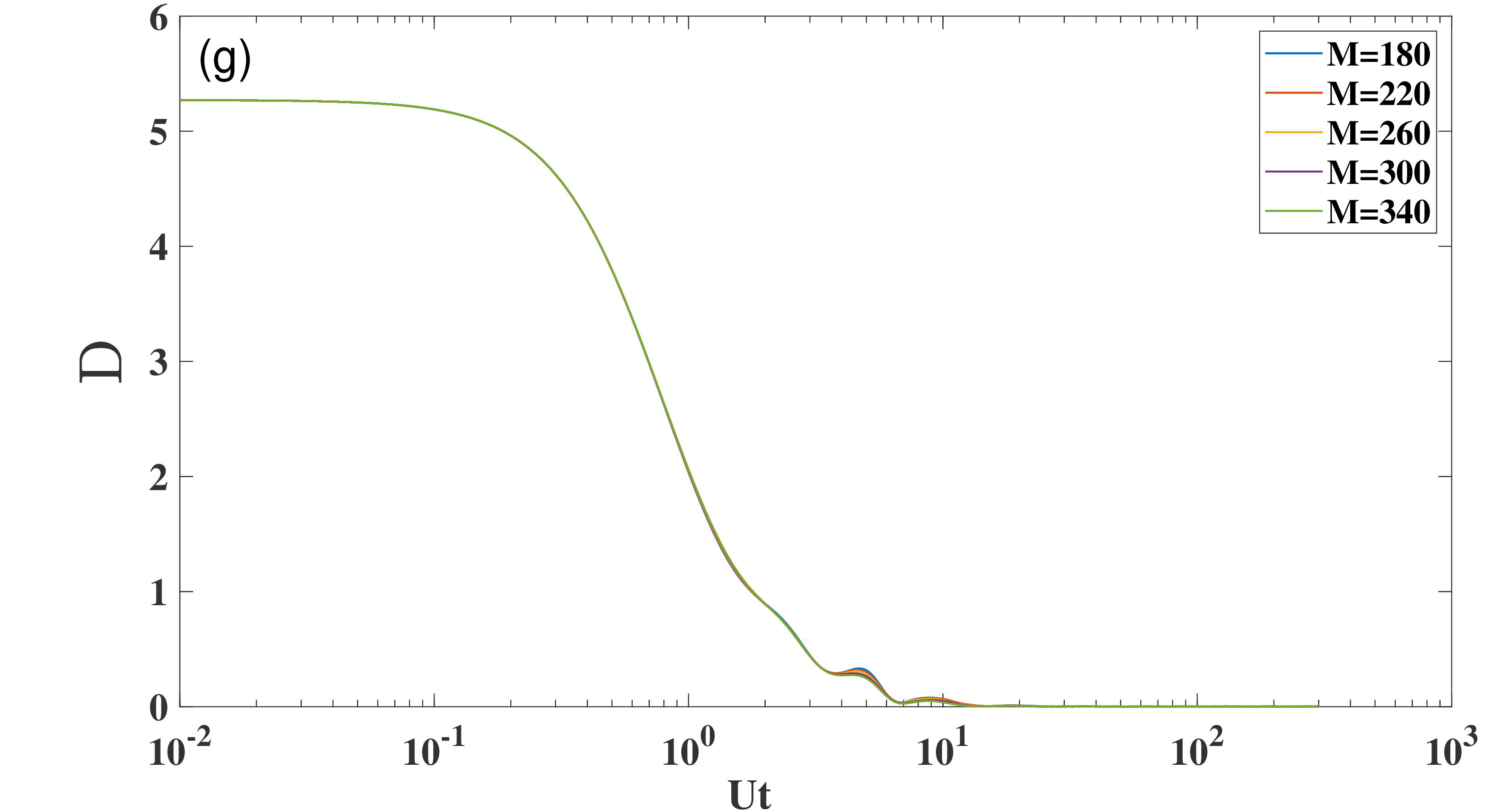}
  \\
  \caption{The population dynamics 
  on odd sites: (a) $i=1$, (b) $i=3$, (c) $i=5$, (d) $i=7$, (e) $i=9$ and (f) $i=11$. For each site, different basis sizes $M$ ranging from 180 to 340 are employed to verify the stability of the results. (g): The dynamics of the sum of the deviations from the equilibrium value, $D(t)=\sum_{i=1}^L |\langle \hat n_i\rangle(t)-1|$. The number of sites is 11, the same as the average particle number. The hopping strength was set to be $J=0.25U$.}\label{fig:circle}
\end{figure}

In the following, we investigate the convergence of the multi-configuration 
method for the case of an average of eleven particles distributed over 
eleven sites in the case of the clean Hamiltonian given in Eq.\ (1).
For a fixed particle number, given by a mean average population of 
eleven, i.e. $N=L=11$, the dimension of the Hilbert space 
(Fock space) is given by ${{L+N-1}\choose{N}}=352716$.  
Even larger spaces emerge for higher particle numbers of, e.g., $N=20$, leading to a Hilbert space dimension of
${{30}\choose{20}}$, which amounts to more than 30 million. For the multi-mode
Glauber state initial condition, one would have to consider the direct 
sum of those variable particle number Hilbert spaces, which makes 
Fock-space calculations impracticable for long-time evolution in the 
present case.

For random initial conditions (the values depicted in Fig.\  \ref{fig:circle}
at the initial points in time) and in the case of 
hopping strength of $J=0.25~U$, a convergence study is shown in 
Fig.\ \ref{fig:circle}. For the figure, we chose to depict results
for the odd sites and our multi-configuration variational approach shows 
that, even though there are tiny fluctuations in the convergence behavior of 
the individual site populations,  the overall behavior and, in particular, 
an observable of interest 
(the sum of  deviations from the thermal value), 
are well converged. Most importantly, the number of
{\it time-dependent} basis function, needed for convergence, is on 
the order of 10$^2$ and  thus several orders of magnitude less than in 
the case of {\it time-independent} Fock states.

\section{Mean site populations in the presence of disorder}

In this section, we perturbatively evaluate how the thermal averages of the individual site populations behave in the presence of weak onsite disorder.
We restrict this analysis to the strongly interacting regime where hopping between adjacent sites of the lattice is relatively weak.
Each lattice site can thus be considered to represent an individual thermodynamic system that is weakly coupled to an effective heat and particle reservoir constituted by the other sites, the latter being characterized by a global temperature $T = 1/\beta$ and a global chemical potential $\mu$.

In the framework of the grand-canonical ensemble, we can describe the thermodynamic system associated with site $l$ through the statistical operator
\begin{equation}
  \hat{\rho}_l = \frac{1}{Y_l} e^{-\beta[\hat{H}_l - \mu \hat{n}_l]}
\end{equation}
where $\hat{n}_l$ is the population operator on site $l$ and $\hat{H}_l$ represents its local energy
\begin{equation}
  \hat{H}_l = \frac{U}{2} \hat{n}_l(\hat{n}_l - 1) + \epsilon_l \hat{n}_l
\end{equation}
with $U$ the (site-independent) onsite interaction and $\epsilon_l$ the (site-dependent) onsite disorder potential.
The partition function of this grand-canonical ensemble is then given by
\begin{equation}
  Y_l = \sum_{n=0}^\infty e^{-\beta[ n ( n - 1 ) U /2 + \epsilon_l n - \mu n]} \,.
  \label{eq:partition}
\end{equation}
and allows one to derive the mean values of the population and of the energy on site $l$ according to
\begin{eqnarray}
  \bar{n}_l & = & \frac{1}{\beta} \frac{\partial}{\partial \mu} \ln Y_l \,, \label{eq:nl} \\
  \bar{E}_l & = & N \mu - \frac{\partial}{\partial \beta} \ln Y_l \,. \label{eq:Hl}
\end{eqnarray}

In the absence of disorder, Eqs.~\eqref{eq:nl} and \eqref{eq:Hl} can be solved, through numerical root searching, to determine the parameters $\beta$ and $\mu$ as a function of the total population and the total energy of the system.
The latter quantities are well known as they can be straightforwardly inferred from the initial state of the system, using the fact that the total energy and the total population of the system are coonstants of motion (and identifying the mean value of the system's total energy approximately with $\sum_l \bar{E}_l$ for weak hopping).
Considering a coherent state that is centered in phase space about the bosonic field amplitudes $\psi_l$, we obtain
\begin{eqnarray}
  \bar{n}_l = \bar{n} & = & \frac{1}{L} \sum_{l=1}^L |\psi_l|^2 \,, \\
  \bar{E}_l = \bar{E} & = & \frac{1}{L} \sum_{l=1}^L \left( \frac{U}{2} |\psi_l|^4 - J (\psi_l^* \psi_{l-1}+ \psi_{l-1}^* \psi_l) \right) \,,
\end{eqnarray}
where we account here also for the hopping contribution (identifying $\psi_0 \equiv \psi_L$).

Let us now consider the presence of nonvanishing but perturbatively small onsite energies $\epsilon_l$.
Without loss of generality, we assume
\begin{equation}
  \sum_{l=1}^L \epsilon_l = 0 \,, \label{eq:disnorm}
\end{equation}
given the fact that global energy shifts, affecting all sites identically, do not have an impact on the mean populations (and can, within Eqs.~\eqref{eq:nl} and \eqref{eq:Hl}, be simply absorbed by a shift of $\mu$).
The presence of this disorder gives rise to a perturbative shift of the system's total energy, thus entailing similarly perturbative shifts in $\mu$ and $\beta$, but cannot affect the total population.
However, the thermal average $\bar{n}_l$ of the local site populations is no longer uniform.
Using the constraint \eqref{eq:disnorm}, we can obtain it by Taylor expanding the expression \eqref{eq:nl} to linear order in $\epsilon_l$ at fixed values for $\mu$ and $\beta$.
This yields
\begin{equation}
  \bar{n}_l = \bar{n} - \epsilon_l \left.\frac{\partial \bar{n}}{\partial \mu}\right|_\beta + O(\epsilon_l^2) \label{eq:Ilin}
\end{equation}
where we obtain from Eq.~\eqref{eq:partition}
\begin{equation}
  \left.\frac{\partial \bar{n}}{\partial \mu}\right|_\beta = \beta \left( \overline{n^2} - \bar{n}^2 \right)
\end{equation}
with $\overline{n^2} = \mathrm{Tr}[\hat{\rho}\hat{n}_l^2]$ evaluated for $\epsilon_l = 0$.
Inserting this expression into Eq.~\eqref{eq:Ilin} confirms the general statistical intuition that for positive temperatures, $\beta > 0$, sites with lower onsite energies should have higher mean populations than sites with higher energies.

\section{Population dynamics in the regime of weak interactions}

As described in the main text, Bose-Hubbard systems are special in the sense that onsite populations in the limit of fully decoupled sites are precisely the quantized actions for the corresponding classical Gross-Pitaevskii lattice. Population dynamics measured at small $J/U$ fractions are, therefore, experimental realizations of perturbation theory in action-angle coordinates. 

For a perfectly harmonic lattice, however, the populations are not related to the classical action variables, which in this regime are given by the lattice's normal modes. Nevertheless, since the regime of strong interactions is simultaneously an integrable and a classical limit of the Gross-Pitaevskii lattice, one has to recover quantum-classical equivalence as $J/U \to \infty$ for any observable, including onsite populations. Thus, although these form a ``bad'' set of observables in the strongly interacting regime in what touches the characterization of thermalization, they are as good a set as any in order to verify that our quantum and classical methods are correct and closely match as $J/U$ is increased.  

\begin{figure}
    \centering
    \includegraphics[width=\linewidth]{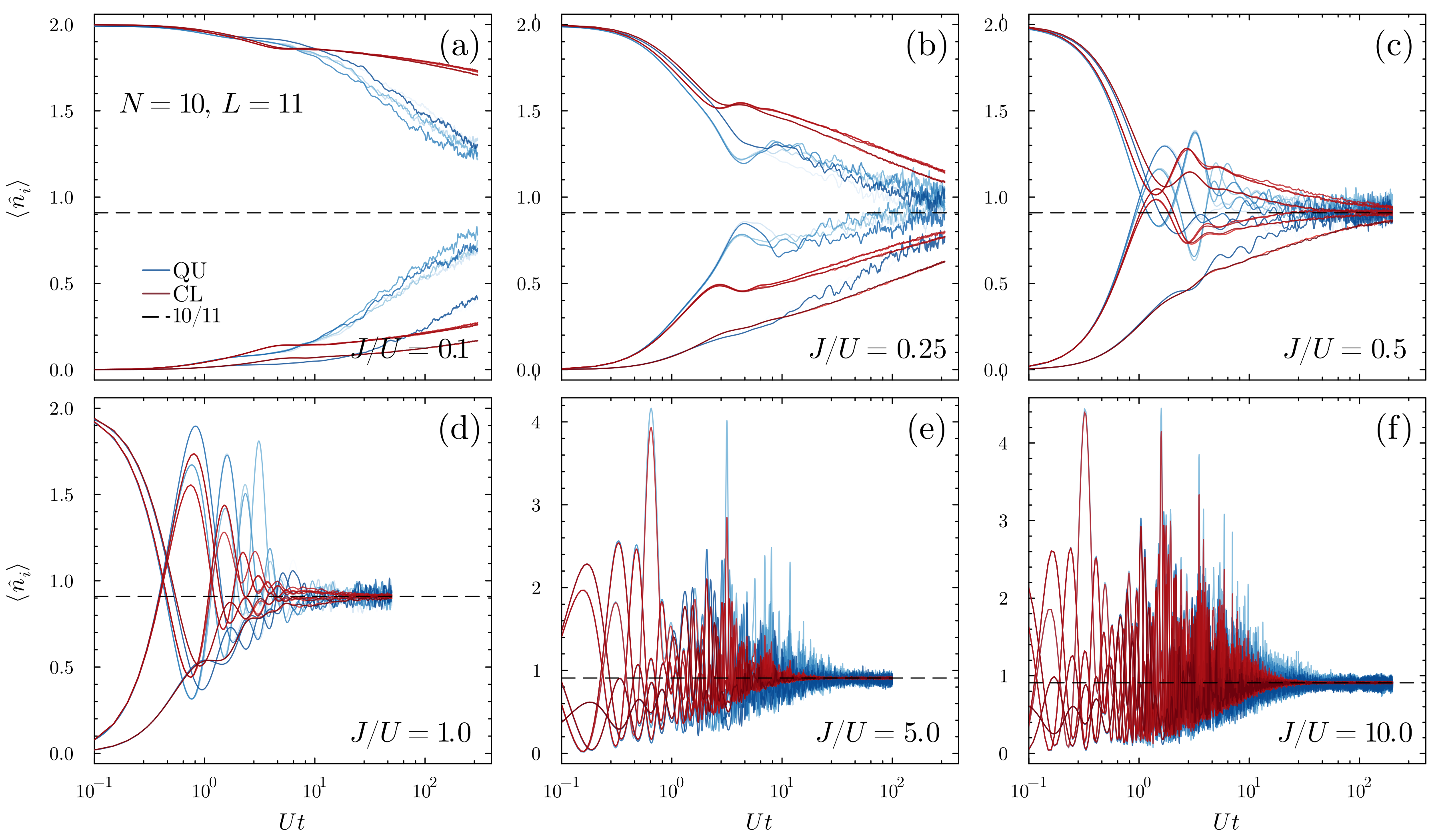}
    \caption{Onsite population dynamics computed classically and quantum mechanically for an initial staggered state $|0,2,0,2,0,2,0,2,0,2,0\rangle$ as a function of increasing $J/U$, \emph{i.e.}~moving from the strong to the weak interaction regime, with $U{=}1$.}
    \label{fig:10_11_pops}
\end{figure}

\begin{figure}
    \centering
    \includegraphics[width=\linewidth]{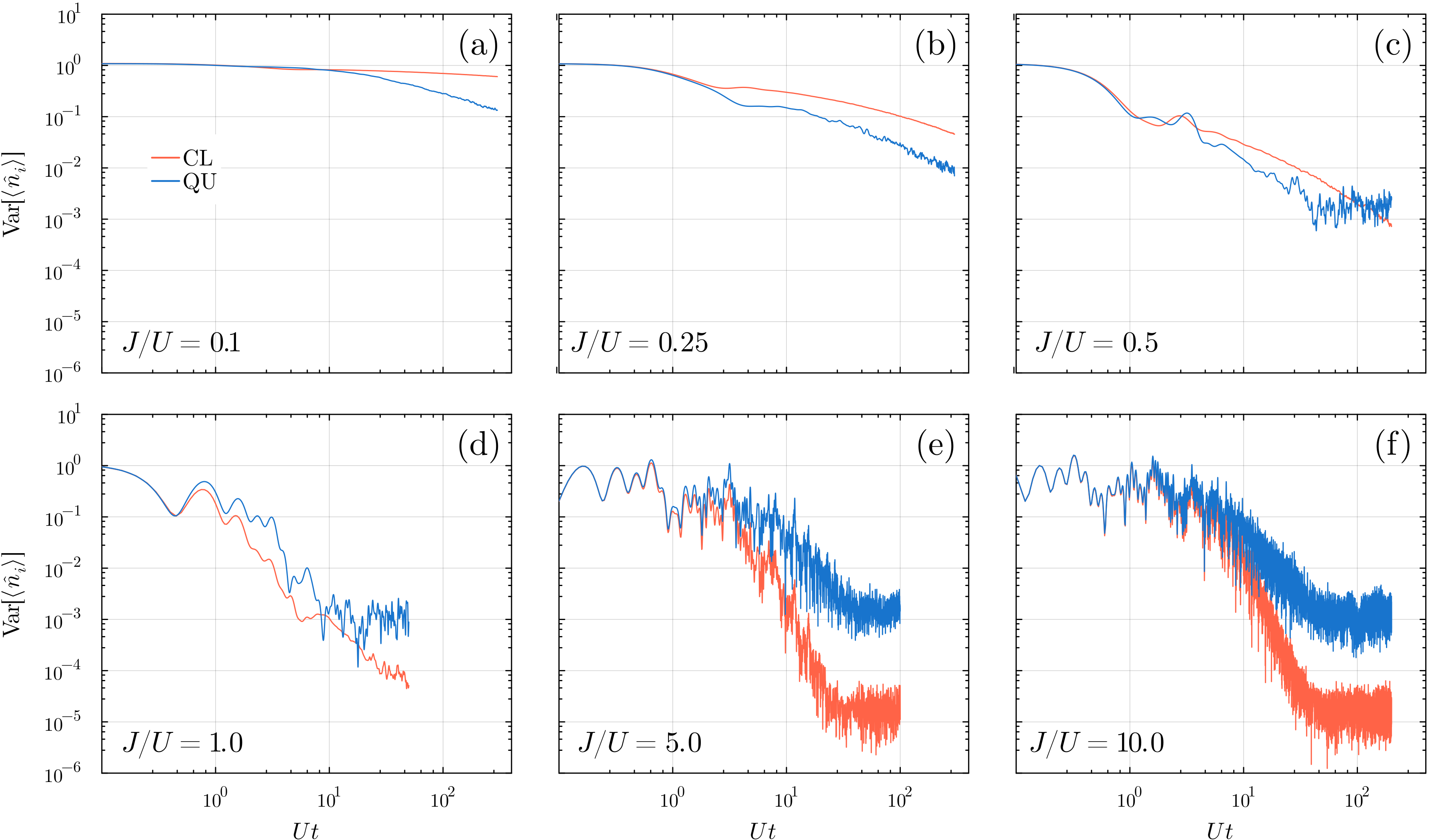}
    \caption{Variances with respect to thermal value obtained from the data in Fig.~\ref{fig:10_11_pops}.}
    \label{fig:10_11_vars}
\end{figure}

In Fig.~\ref{fig:10_11_pops} we display the population dynamics of an initial staggered state (similar to the one in Fig.~3 of the main text) in a system with $N{=}10$ and $L{=}11$ as it transitions from strong to the weak interaction regime. As clearly seen from moving rightward from Fig.~\ref{fig:10_11_pops}(a), QET fades as $J/U$ is increased, with quantum and classical onsite populations equilibrating at roughly the same time stating from Fig.~\ref{fig:10_11_pops}(d). As the interaction strength increases in Fig.~\ref{fig:10_11_pops}(e) and, especially, Fig.~\ref{fig:10_11_pops}(f), we see that quantum and classical populations fall essentially on top of each other until equilibration is reached. As expected, the finiteness of an effective Planck constant $\hbar_\mathrm{eff}{=}1/N$ means that quantum onsite populations oscillate around the thermal average with higher amplitude than the classical ones. The approach towards thermal equilibrium can also be tracked in the variances of Fig.~\ref{fig:10_11_vars}, with the first panels displaying once again the standard QET characteristics. Note that in the last two panels of Fig.~\ref{fig:10_11_vars} ``equilibrium'' means that the methods have each reached their numerical accuracies, which are around $10^{-3}$ and $10^{-5}$ for the quantum and classical computations, respectively. Since these correspond to $0.1\%$ and $0.001\%$ of the equilibrium value it is clear that these findings, together with the convergence tests of Sec.~\ref{sec:var}, attest for the accuracy of our variational method used to obtain quantum populations and shows once again that QET is a robust dynamical phenomenon. 

\section{EFFECTIVE COUPLING BETWEEN NEAR-RESONANT STATES}
\begin{figure}[!htbp]
  \centering
  \includegraphics[width=6in]{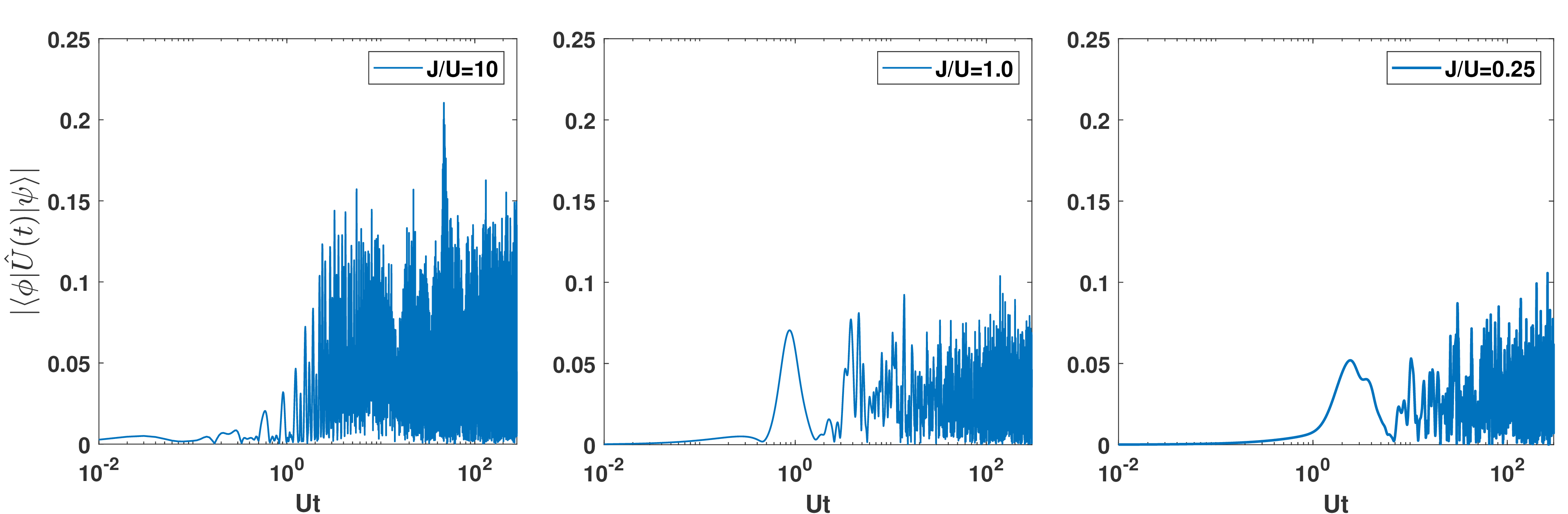}
  \caption{Absolute value of the transition amplitude, $|\langle\phi|\hat U(t)|\psi\rangle|$, between two Glauber coherent states $|\psi\rangle = |0,2,0,2,0,2,0\rangle$ and $|\phi\rangle = |2,0,2,0,2,0,0\rangle$ with different hopping strengths: (a) $J/U=10.0$, (b) $J/U=1.0$ and (c) $J/U=0.25$.}\label{Fig:transition_well3}
\end{figure}

\begin{figure}[!htbp]
  \centering
  \includegraphics[width=6in]{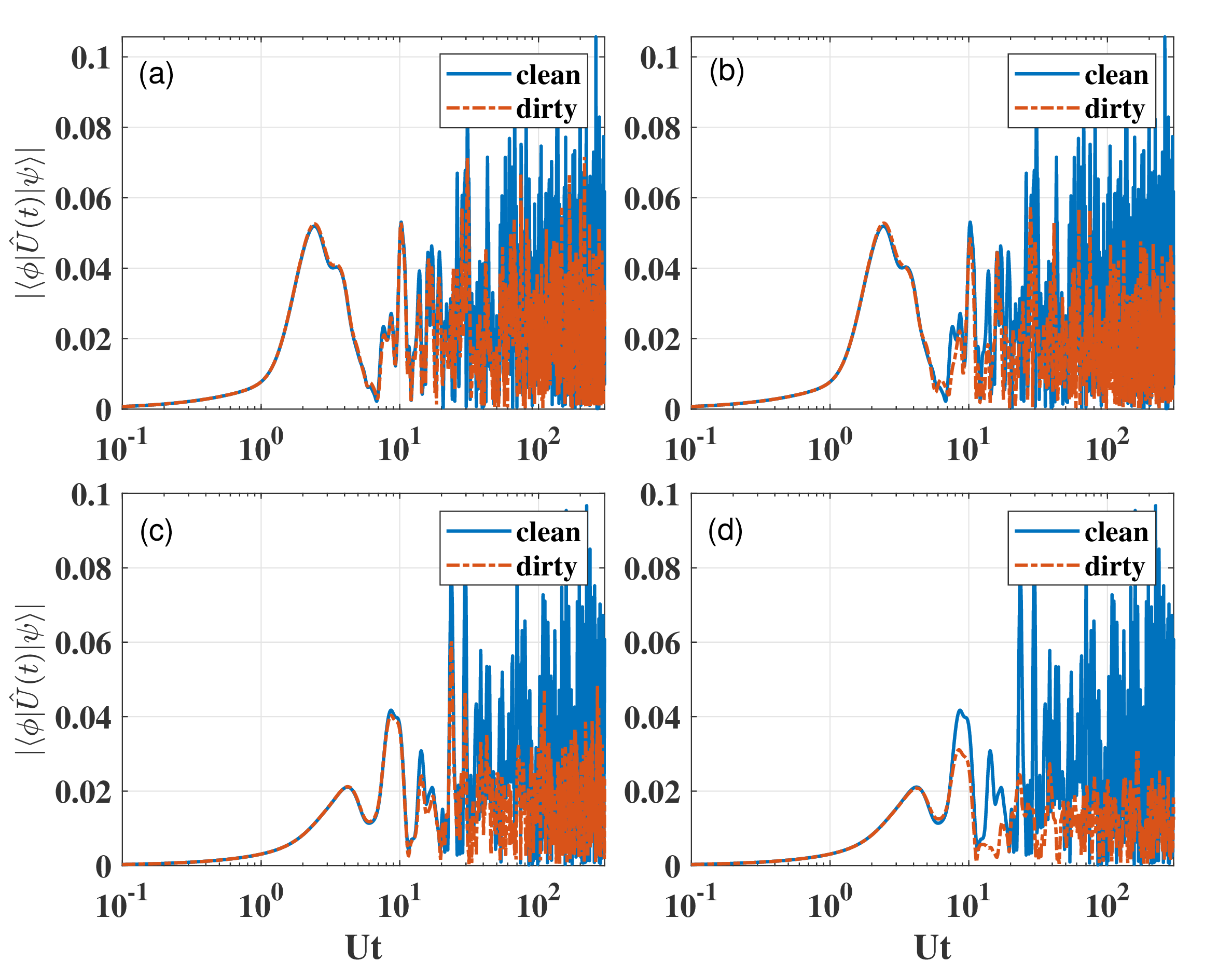}
  
  \caption{Absolute value of the transition amplitude, $|\langle\phi|\hat U(t)|\psi\rangle|$, between two Glauber coherent states $|\psi\rangle = |0,2,0,2,0,2,0\rangle$ and $|\phi\rangle = |2,0,2,0,2,0,0\rangle$. Panels (a) and (b) use the same parameters as in Fig.\ 3(a) and Fig.\ 3(b) of the main text with the hopping strength $J=0.25U$. The blue solid lines show the transition amplitudes for a disorder-free system whereas random onsite energies, uniformly sampled within $[-0.05U,0.05U]$ in (a) and within $[-0.25U,0.25U]$ in (b),  are considered for the red dashed lines. In (c) and (d), the hopping strength is reduced to be $0.1U$, and the settings of disorder strength follow those used in (a) and (b), respectively.}\label{Fig:transition}
\end{figure}

To support the conjecture proposed in the main text, namely that QET arises from approximate resonances between the initial state and a variety of other states obtained from it through permutations of the site occupancies, we select two representative Glauber coherent states $|\psi\rangle$ and $|\phi\rangle$ and calculate their quantum transition amplitude $\langle\phi|\hat U(t)|\psi\rangle$ with $\hat U(t)={\rm e}^{-{\rm i}\hat H t/\hbar}$. Specifically, in Fig.\ \ref{Fig:transition_well3} and Fig.\ \ref{Fig:transition}, $|\psi\rangle$ and $|\phi\rangle$ are chosen as $|\psi\rangle=|0,2,0,2,0,2,0\rangle$ and $|\phi\rangle=|2,0,2,0,2,0,0\rangle$. These two states have vanishing overlap with each other but share the same energy in the absence of disorder. In Fig.\ \ref{Fig:transition_well3}, we compare the dynamics of the transition amplitude $|\langle\phi|\hat U(t)|\psi\rangle|$ for different hopping strengths. We find that a larger hopping strength (or equivalently, a weaker interaction strength) enhances the transition amplitude between the two Glauber coherent states. However, as shown in panels (b) and (c) of Fig.\ \ref{Fig:transition_well3}, the transition amplitude remains significant for weaker hopping strengths. In fact, due to the lack of proper dynamical tunneling barriers between these two states in the weak hopping regime (owing to the fact that the system is fully chaotic), they can be coupled rather effectively by quantum dynamics through multiple channels mediated by numerous intermediate states. Such effective coupling will facilitate the exploration of the evolved states over different phase-space regions and thereby enhance the process of thermalization. Thus, in the regime where quantum and classical systems thermalize on comparable timescales, equipartition in site populations happens faster and correlates with an increased magnitude of transition amplitudes. Once QET starts taking place, equipartition slows down and transition amplitudes also become smaller.

When onsite disorder is included in the weak hopping regime as shown in Fig.\ \ref{Fig:transition}, exact degeneracy between the levels associated with $|\psi\rangle$ and $|\phi\rangle$ is broken, leading to a reduction of the effective coupling. Intriguingly, the decreasing coupling induced by the disorder becomes apparent only for $t\geq10/U$, which coincides with the onset of thermalization suppression, as shown in Fig.\ 3(a) and Fig.\ 3(b) for different disorder strength. 
\\
\\
\\
\\
\\
\bibliography{bib}